\documentclass[11pt]{article}
\usepackage[scale=.67,centering]{geometry}
\usepackage{amssymb,amsmath}
\usepackage{latexsym}
\usepackage{natbib}
\usepackage{fleqn}
\usepackage{smallsec,nustyle2}
\usepackage[usenames]{color}
\renewcommand{\baselinestretch}{1.23}
\usepackage{times}

\title{Notes on random reals%
\thanks{Contact:
osherson@princeton.edu, weinstein@cis.upenn.edu.
}}

\author{%
Daniel Osherson \\ Princeton University \and
Scott Weinstein \\ University of Pennsylvania}

\newcommand{\PROB}{\mbox{\textsl{Pr}}}
\newcommand{\CONCAT}{\star}
\newcommand{\N}{\mbox{$\mathbb{N}$}}
\newcommand{\BINARY}{\mbox{\textsf{B}}}

\newcommand{\UPC}[1]{ #1^{\sharp} }

\newcommand{\MID}{\mbox{$\ :\ $}}
\newcommand{\DOMAIN}{\mbox{\textsl{domain}}}
\newcommand{\RANGE}{\mbox{\textsl{range}}}
\newcommand{\QED}{\ensuremath{\hfill \Box}}

\newcommand{\OB}{\textbf}
\newcommand{\KC}{\mbox{${\cal C}$}}
\newcommand{\PCS}{\mbox{\footnotesize{\mbox{${\cal K}$}}}}
\newcommand{\PC}{\mbox{${\cal K}$}}
\newcommand{\PROOF}{\noindent\textsl{Proof:}\ \,}
\newcommand{\SPOOF}[1]{\noindent\textsl{Proof of #1:}\ \,}
\newcommand{\TM}{\textsf{TM}}
\newcommand{\BN}[1]{\ensuremath{\tau(#1)}}
\newcommand{\PF}{prefix-free}
\newcommand{\PFC}{Prefix-free}

\newtheorem{prop}[equation]{Proposition: }
\newtheorem{cor}[equation]{Corollary: }
\newtheorem{thm}[equation]{Theorem: }
\newtheorem{defn}[equation]{Definition: }
\newtheorem{conv}[equation]{Convention: }
\newtheorem{fact}[equation]{Fact: }
\newtheorem{dispar}[equation]{}
\newtheorem{claim}[equation]{Claim: }
\newtheorem{lem}[equation]{Lemma: }
\newtheorem{ex}[equation]{Example: }

\newcommand{\LENGTH}[1]{ |#1| }

\newcommand{\ML}{Martin-L\"of}
\newcommand{\DOWARR}{\!\downarrow\,}
\newcommand{\UPARR}{\!\uparrow\,}

\begin{document}
\maketitle

The theory of random real numbers is exceedingly well-developed,
and fascinating from many points of view. It is also quite
challenging mathematically. The present notes are intended
as no more than a gateway to the larger theory. They review
just the most elementary part of the theory (bearing on
Kolmogorov- and \ML-randomness). We hope that the simple
arguments presented here will encourage the enterprising
student to examine richer treatments of the subject
available elsewhere, notably, in
\citet{DH}.\footnote{One small conceptual
contribution to the theory is offered in \citet{OW06}.}
Comments and corrections to the notes are, of course,
welcome.

\section{Notation and other preliminaries}\label{Notsec}

Let $\N = \{0, 1, 2, \ldots\}$.
By a \OB{sequence} we
mean an infinite sequence ordered like \N. Given $n\in \N$ and
infinite sequence $x$, we use $x(n)$ to denote the $n$th
member of $x$, and $x[n]$ to denote the initial finite
sequence of length $n$ in $x$. So $x[0]$ is the empty sequence.
An infinite sequence over $\{0,1\}$ is called a \OB{real}. Let \BINARY\ be the set of
\textit{finite} binary strings.
The concatenation of $b,c\in\BINARY$
is denoted $b\CONCAT c$. For $n\in\N$, $1^n$ denotes $n$ $1$'s in a row.
The length of $b\in\BINARY$ is denoted $\LENGTH{b}$.
For $b\in\BINARY$ and infinite binary sequence $x$
we write $b \subset x,$ just in case there is an $n\in\N$ with $b = x[n].$
Similarly, for $a,b\in\BINARY$ we write $b \subseteq a,$ just in case $b$ is an initial segment of $a.$
For $b\in\BINARY$, $O(b)$ denotes the set of reals that begin with $b$.
For $B\subseteq\BINARY$, $O(B) = \{O(b)\MID b\in B\}$. All uses of $\log$ are base $2$.

We fix an effective enumeration \BN{m} of \BINARY, $m\in\N$.
Members of \BINARY\ are enumerated
according to length with ties broken lexicographically.

\begin{lem}\label{cblem}
For all $m\in\N$, $| \BN{m} | \le \log(m+1)$.
\end{lem}

\begin{defn}\label{closed}
Let $Q\subseteq\BINARY$ be given. We say that $Q$ is \OB{closed under subsequences} just
in case for all $b\in Q$, $a\in Q$ for every $a\subseteq b$.
\end{defn}
For proof of the the following well-known result see \citet[pp.\ 323-324]{BBJ}.
\begin{lem}\label{KonigLemDan}(K\"{o}nig)
Let $Q\subseteq\BINARY$ be infinite and closed under subsequences. Then there is a
real $x$ such that $x[n]\in Q$ for all $n$.
\end{lem}

We sometimes identify a real $x$ (as understood here as an infinite sequence
over $\{0, 1\}$) with a
real number in the interval $[0,1]$ via the map $I$ that sends $x$
to $\sum_{i=0}^\infty x(i)\cdot2^{-(i+1)}.$ Similarly, we may identify
the finite initial segment $x[n]$ with the rational number
$\sum_{i=0}^{n-1} x(i)\cdot2^{-(i+1)}$. Any such finite sum yields a rational
number called \OB{dyadic}. Observe that $I^{-1}$ is well-defined for all reals in
$[0,1]$ aside from the positive dyadic rationals. Indeed, each dyadic
rational besides $0$ is the $I$ image of exactly two infinite binary
sequences. One of the sequences ends in a tail of $0$'s, the other
in a tail of $1$'s. For example, the real number $\frac18$ equals
both $I(111000\cdots)$ and $I(110111\cdots)$. When mapping real numbers
into real binary sequences, we must therefore choose between the two kinds
of tails. The theory below can be developed on the basis of either choice.
We prefer the latter.
\begin{conv}\label{dyadicConv}
For every positive, dyadic rational number $r$, we take its representation
as a sequence over $\{0, 1\}$ to
end in an infinite sequence of $1$'s.
\end{conv}
The convention resolves all ambiguity; every other real number
has a unique inverse image under $I$.

It is tempting to introduce notation that distinguishes between ``reals''
in the sense of sequences over $\{0, 1\}$ from ``reals'' in the sense
of numbers (points on the real line). We bow to custom, however, and rely
on context to clarify whether ``real'' is used in one sense or the other.
Note that whenever a statement involves an inequality (or weak inequality)
then the map $I$ is used implicitly. Similarly, when a real number $x$
is juxtaposed with $[n]$ to form $x[n]$ then we have first implicitly applied
$I^{-1}$ to $x$ to produce a sequence over $\{0, 1\}$.
The following lemma makes use of these conventions, and
codifies a few well-known facts that we
will refer to in later sections.
\begin{lem}\label{geofacts}
\begin{enumerate}
%\item\label{geofactsa}
%\[\sum_{i = 1}^{n}\left(\frac12\right)^{i} = 1 - \left(\frac12\right)^{n}.\]
\item\label{geofactsb}
\[\sum_{i = n+1}^{\infty}\left(\frac12\right)^{i} =
\left(\frac12\right)^{n}.\]
\item\label{chaitin2b} For all reals $x$ and all $n \in \N,$
$x[n] \le x \le x[n] + 2^{-n}$.
\item\label{oddconvention} For every real number $x \in (0,1],$ $x[n] < x.$
\item\label{indFact} For all $n\in\N$,
$\sum_{i < n}2^{i} = 2^n - 1$.
\end{enumerate}
\end{lem}
\PROOF %
Fact (\ref{geofactsb}) is an immediate consequence of
the following well-known identity by setting $r = 1/2.$
%\[(1-r)\cdot \sum_{i=1}^n r^i = r - r^{n+1},\] and
\[\text{For every real
number $0\le r < 1$ and $n \in \N,$}\quad (1-r)\cdot \sum_{i=n}^{\infty}  r^i = r^n.\]

Fact (\ref{chaitin2b}) follows immediately from (\ref{geofactsb}).

Fact (\ref{oddconvention}) follows from Convention \ref{dyadicConv} concerning
the identification of real numbers in $(0,1]$ with infinite binary
sequences which are {\em not} eventually constantly $0.$

Fact (\ref{indFact}) is easily proved by induction.
\QED

\section{Kolmogorov complexity}

Let $W_i$ index the computably
enumerable subsets of \BINARY\ (instead of indexing the computably
enumerable subsets of $N$, which is more usual). The indexes
on the $W_i$ are qualified as ``r.e.'' (recursively enumerable).

We use \TM\ to abbreviate ``Turing Machine.''
Members of \BINARY\ are conceived to be the inputs
and outputs of \TM s.
Let $M$ be a \TM, and let $a,b\in\BINARY$ be given. We
write $M(a) = b$ just in case $M$ started with $a$ on its tape halts
with $b$ on its tape.
We write $\TM(b)\DOWARR$ to signify that
$\TM(b)$ is so defined, and $\TM(b)\UPARR$ to signify
that it is undefined.
Via a fixed, effective bijection between
\BINARY\ and the set of \TM s,
\TM s are themselves taken to be members of \BINARY.
\TM s thus inherit the ordering imposed above on \BINARY.

We write $\KC_M(b)$ to be the length of a
shortest $c\in\BINARY$ such that $M(c) = b$; $\KC_M(b) = \infty$ if
no $c\in\BINARY$ is such that $M(c) = b$; this number is known as
the ``plain Kolmogorov complexity'' of $b$ relative to $M$.

\begin{defn}\label{Udef}
\TM\ $M$ is called \OB{universal}
just in case for every \TM\ $L$ there is $k\in\N$ such that
$\KC_M(b) \le \KC_L(b) + k$ for all $b\in\BINARY$.
\end{defn}
Note that the definition implies that
$\KC_M(b)$ is finite for every $b\in\BINARY$ if $M$ is universal.
It's also worth observing that ``universal'' in the sense of
Definition \ref{Udef} does not yield the same set of machines
as ``universal'' in Turing's original sense.

\begin{lem}\label{UMexist}
Universal \TM s exist.
\end{lem}
\PROOF Let $A$ be a lexicographical ordering of all \TM s.
It follows immediately from the existence of universal machines
in Turing's sense that there is a \TM\ $U$
such that for all $n\in\N$ and $d\in\BINARY$, $U(1^n0d) = L(d)$,
where $L = A(n)$. To verify that
$U$ is universal, let \TM\ $L$ be given, and let $n$ be such that
$L = A(n)$. Then for all $b\in\BINARY$, $\KC_U(b) \le \KC_L(b) + n + 1$. \QED

In light of the lemma, we fix a universal \TM\ $U$, and we write
$\KC(b)$ in place of $\KC_U(b)$. Let $L$ be the \TM\ that halts
immediately, making no changes to its tape. Then for all $b\in\BINARY$,
$\KC_L(b) = \LENGTH{b}$. Because $U$ is universal there is $m\in\N$ such
that $\KC(b) \le \KC_L(b) + m$. It follows at once that:
\begin{lem}\label{evFact1}
There is $m\in N$ such that for every $b\in\BINARY$,
$\KC(b) \le \LENGTH{b} + m$.
\end{lem}

\section{A fact about short instructions}

An input to $U$ can be conceived as instructions for producing an output.
Inputs that are the shortest possible for producing their output are called
\textit{short}. Officially:

\begin{defn}\label{shortDef}
Call $p\in\BINARY$ \OB{short} if $U(p)\DOWARR$, and for all
$q\in\BINARY$, if $U(p) = U(q)$ then $\LENGTH{p} \le \LENGTH{q}$.
\end{defn}

\noindent Equivalently:
\begin{dispar}\label{shortEq}
$p\in\BINARY$ is short iff
$\KC(U(p)) = \LENGTH{p}$.
\end{dispar}

\begin{prop}\label{shortFact}
There is no effective enumeration of an infinite number
of short members of \BINARY.
\end{prop}

\def\NPLUS{\ensuremath{\N^+}}
\PROOF
We follow \citet[p.\ 121]{Li}.
For a contradiction,
let $p_i$, $i\in \N$ be an enumeration of infinitely many
short members of \BINARY. Then by \ref{shortEq},
$\{\KC(p_i)\MID i\in \N\}$ is unbounded. Therefore,
the following function, $g:\BINARY\rightarrow \BINARY$, is total
and computable.
\[\mbox{For all $m\in \N$, } g(\BN{m}) = p_i\mbox{ where $i$ is least such that }\KC(p_i) \ge m.\]
By the definition of $g$, we have:
\begin{dispar}\label{shortev1}
for all $m\in \N$,
$\KC(g(\BN{m})) \ge m$.
\end{dispar}
Suppose that \TM\ $L$ computes $g$. Then
by the universality of the reference machine $U$ there is $k\in \N$ such that:
\begin{dispar}\label{shortev2}
for all $m\in \N$, $\KC(g(\BN{m})) \le \KC_L(g(\BN{m})) + k$.
\end{dispar}
Also, since the string $\BN{m}$ causes $L$ to produce $g(\BN{m})$, and by Lemma
\ref{cblem}:
\begin{dispar}\label{shortev3}
for all $m\in \N$, $\KC_L(g(\BN{m})) \le \LENGTH{\BN{m}}\le \log(m+1) + 1$.
\end{dispar}
{}From \ref{shortev1} and \ref{shortev2}:
\begin{dispar}\label{shortev4}
for all $m\in \N$, $m \le \KC_L(g(\BN{m})) + k$.
\end{dispar}
And from \ref{shortev3} and \ref{shortev4}:
\begin{quote}
for all $m\in \N$, $m \le \log(m+1) + k + 1$
\end{quote}
which is false no matter which $k\in \N$ is chosen. \QED

\section{Failure of a plausible account of randomness}

A promising idea is to qualify real $x$ as random just in case
cofinitely many of $x$'s initial segments have high complexity.
In this section we specify this idea and show how it comes to grief.

\begin{defn}\label{incomDef}
Call $b \in \BINARY$ \OB{incompressible} just in case
$\KC(b) \ge \LENGTH{b}$.
\end{defn}
Since there are $2^n$ binary strings of
length $n$ and only $\sum_{i < n}2^{i} = 2^n - 1$ inputs to $U$ of length less than $n$
[see \ref{geofacts}\ref{indFact}],
it follows that:

\begin{lem}\label{incomExist}
For every $n \in \N$ there are incompressible $b\in\BINARY$ with
$\LENGTH{b} = n$.
\end{lem}

\begin{defn}\label{incomrDef}
Call a real $x$ \OB{incompressible almost always} just in case
\[\{n\MID x[n]\mbox{ is incompressible}\}\]is cofinite.
\end{defn}
We might hope that the set of reals that are incompressible almost always
is rich and numerous, but
it turns out to be empty! We'll derive this surprising fact
as a corollary to the following
proposition.

\begin{prop}\label{doesNotWork}
For all real $x$ and all $k\in \N$ there is $n\in\N$ such that
$\KC(x[n]) < n - k$.
\iffalse
For no real $x$ is there $k\in\N$ such that
for all $n\in N$, $\KC(x[n]) \ge n - k$.
\fi
\end{prop}

\noindent To prove the proposition, we start with a lemma.

\begin{lem}\label{lengthened}
Let total recursive function $f:\BINARY\rightarrow\BINARY$ be given. Then
there is $k\in\N$ such that for all $b\in\BINARY$,
$\KC(f(b)) < \KC(b) + k$.
\end{lem}

\PROOF Recall that $U$ is our fixed universal \TM, and let \TM\ $L$ be such that:
\begin{dispar}\label{lengthened1}
for all $a\in\BINARY$, $L(a) = f(U(a))$.
\end{dispar}
By Definition \ref{Udef}, let $k\in\N$ be such that:
\begin{dispar}\label{lengthened2}
for all $c\in\BINARY$, $\KC(c) < \KC_L(c) + k$.
\end{dispar}
Let $b\in\BINARY$ be given, and let $a\in\BINARY$ be of shortest length
with $U(a) = b$. Hence:
\begin{dispar}\label{lengthened3}
$\KC(b) = \LENGTH{a}$.
\end{dispar}
By \ref{lengthened1}, $L(a) = f(b)$, hence:
\begin{dispar}\label{lengthened5}
$\KC_L(f(b)) \le \LENGTH{a}$.
\end{dispar}
By \ref{lengthened2} and \ref{lengthened5},
$\KC(f(b)) < \KC_L(f(b)) + k < \LENGTH{a} + k$, so by \ref{lengthened3},
$\KC(f(b)) < \KC(b) + k$. \QED

\SPOOF{Proposition \ref{doesNotWork}} Recall that $\BN{\cdot}$ is an effective
bijection between \N\ and \BINARY. Let effective $f:\BINARY\rightarrow\BINARY$ be such that for all
$b\in\BINARY$, $f(b) = \BN{\LENGTH{b}}\CONCAT b$. By Lemma \ref{lengthened} there
is $k_0\in\N$ such that for all $b\in\BINARY$,
$\KC(f(b)) < \KC(b) + k_0$. So by Lemma \ref{evFact1} there is $k_1\in\N$ such that
for all $b\in\BINARY$,
$\KC(f(b)) < \LENGTH{b} + k_1$. Thus:
\begin{dispar}\label{doesNotWorka}
for all $b\in\BINARY$,
$\KC(\BN{\LENGTH{b}}\CONCAT b) < \LENGTH{b} + k_1$.
\end{dispar}

Now let real $x$ and $k\in\N$ be given. To prove the proposition we must exhibit
$n\in\N$ such that:
\begin{dispar}\label{doesNotWorkb}
$\KC(x[n]) < n - k$.
\end{dispar}
Choose $p\in\N$ such that $\BN{p}\subset x$ and $\LENGTH{\BN{p}} > k_1 + k$. (That there
is such a $p$ is obvious.) Let $b\in\BINARY$ be the $p$ bits of $x$ following $\BN{p}$, and let
$n$ be the length of $\BN{p}\CONCAT b$. Thus:
\begin{dispar}\label{doesNotWorkc}
\begin{enumerate}
\item\label{doesNotWorkca}
$\LENGTH{b} = p$
\item\label{doesNotWorkcaa}
$x[n] = \BN{p}\CONCAT b = \BN{\LENGTH{b}}\CONCAT b$
\item\label{doesNotWorkcc}
$\LENGTH{h\BN{\LENGTH{b}}} = \LENGTH{\BN{p}} > k_1 + k$
\item\label{doesNotWorkcb}
$\LENGTH{x[n]} = \LENGTH{\BN{\LENGTH{b}}\CONCAT b} = \LENGTH{\BN{p}} + \LENGTH{b} > k_1 + k + p$
\end{enumerate}
\end{dispar}
By \ref{doesNotWorkc} and \ref{doesNotWorka}:
\[
\KC(x[n]) = \KC(\BN{\LENGTH{b}}\CONCAT b) < \LENGTH{b} + k_1 = p + k_1 = (k_1 + k + p) - k < \LENGTH{x[n]} - k,
\]
which verifies \ref{doesNotWorkb}.
\QED

\begin{cor}\label{incompCor}
No real is incompressible almost always.
\end{cor}

\PROOF Suppose for a contradiction that real $x$ is incompressible almost always. Then
\[k = \sum\{i+\KC(x[i])\MID x[i]\mbox{ is not incompressible }\}\] is
well defined. It follows that for all $n\in\N$,
$\KC(x[n]) \ge n - k$, contradicting Proposition \ref{doesNotWork}.

\QED

\section{No subadditivity for \KC}

The following proposition is meant to deepen the conviction that \KC\ is
not the right measure of complexity for finite sequences. (But we admit
to not understanding why this feature of \KC\ is considered a defect.)

\begin{prop}\label{subAdd1} For every $\ell\in N$,
there are $a,b\in\BINARY$ such that
$C(a\CONCAT b) \ge C(a) + C(b) + \ell$.
\end{prop}

\noindent
To prove the proposition, we start with two lemmas.

\begin{lem}\label{koniglem}
Let $P \subseteq \BINARY$ and suppose that for all real $x$ there
is an $n\in\N$ such that $x[n] \in P.$ Then, there is an $m \in\N$
such that for all real $x$ there is an $n <m$ such that  $x[n] \in P.$
\end{lem}
\SPOOF{Lemma \ref{koniglem}}\ Suppose that
\begin{dispar}\label{hypScott}
for every $m \in \N$ there is an $x$ such that for all $n < m$ $x[n] \not\in P.$
\end{dispar}
Let $Q  = \{a \in \BINARY \mid \forall b ( b \subseteq a \rightarrow b \not\in P\}.$ It follows
from \ref{hypScott} that $Q$ is infinite and closed under subsequences. Therefore, by Lemma \ref{KonigLemDan}
there is a real $x$ such that for all $n \in N$ $x[n]\in Q.$
Since $Q \subseteq \overline{P}$, this
contradicts the hypothesis of \ref{koniglem}. \QED

\begin{defn}\label{kcomDef}
Call $b \in \BINARY$ \OB{$k$-compressible} just in case
$\KC(b) \le \LENGTH{b} - k$.
\end{defn}

\begin{lem}\label{doesNotWorkScott}
For every $k\in\N,$ there is an $m \in\N$ such that for every $a \in\BINARY$,
if $\LENGTH{a} \ge m,$ then there is $d \subset a$ such that $d$ is $k$-compressible.
\end{lem}
\PROOF Fix $k$ and let $P \subseteq \BINARY$ be the collection of $k$-compressible sequences. Proposition \ref{doesNotWork} guarantees that for every real $x$ there is an $n$ such that $x[n]\in P.$ The lemma now follows at once from Lemma \ref{koniglem}. \QED

\SPOOF{Proposition \ref{subAdd1}}\ Fix $\ell\in N$. By Lemma
\ref{evFact1}, choose $m$ so that for every $b\in\BINARY$, $\KC(b)
\le \LENGTH{b} + m$. Let $k = \ell + m.$ By Lemmas \ref{incomExist}
and \ref{doesNotWorkScott}, we may choose an incompressible $a \in
\BINARY$ and $d \subset a$ with $d$ $k$-compressible. Let $b$ be such
that $a = d\CONCAT b.$
Then
\begin{eqnarray*}
\KC(d\CONCAT b) = \KC(a) \ge \LENGTH{a} = \LENGTH{d\CONCAT b} = \LENGTH{d} + \LENGTH{b} \ge \KC(d) + k + \LENGTH{b}
\\
\ge \KC(d) + m + \ell + \KC(b) - m = \KC(d) + \KC(b) + \ell. %\quad \mbox{\QED}
\end{eqnarray*}
\QED

\section{Prefix-free sets}

The defects in plain Kolmogorov complexity lead to an approach
based on ``prefix-free'' subsets of \BINARY.

\begin{defn}\label{prefixFreeTM}
\begin{enumerate}
\item\label{prefixFreeTMa} $S\subseteq\BINARY$ is
\OB{\PF} just in case for all $a,b\in S$, neither
$a\subset b$ nor $b\subset a$.
\item\label{prefixFreeTMb}
A \TM\ $L$ is \OB{\PF}
just in case $\DOMAIN(L)$ is \PF\ [that is,
for all $b \in\BINARY$ and
$c \subset b$, $L(b)\DOWARR$ implies
$L(c)\UPARR$].
\end{enumerate}
\end{defn}
\begin{ex}\label{pfEx1}
$T = \{b\in\BINARY\MID \mbox{ for some }n\in\N, b = 1^n\CONCAT0\CONCAT a\mbox{ with }
\LENGTH{a} = n\}$ is \PF, infinite, and effectively enumerable.
\end{ex}

\begin{lem}\label{irredundancy-lemma}
For every r.e.\ $S\subseteq\BINARY$ there is an r.e.\
$T\subseteq \BINARY$ such that
\begin{enumerate}
\item $O(S) = O(T)$ and
\item $T$ is \PF.
\end{enumerate}
Moreover, an index for $T$ can be found uniformly effectively from
an index for $S$.
\end{lem}

\PROOF Given a recursive enumeration $s_0, s_1, \ldots$ of $S$, $T$ can be
constructed by the following induction which effectively constructs a chain
$T_0 \subseteq T_1 \subseteq \dots \subseteq \BINARY$ with
$T = \cup_i T_i$. (If $S = \emptyset$ then the construction will deliver
an index for $\emptyset$.) \underline{Basis}: $T_0 = \{ s_0\}$.
\underline{Induction step}: Suppose the induction has been completed through
stage $n$. Let $T_n$ be the subset of $T$ already defined. If $s_{n+1}$ is neither
a suffix nor prefix of any $t\in T_n$ then $T_{n+1} = T_n \cup \{ s_{n+1}\}$.
If for some $t\in T_n$,
$s_{n+1}$ extends $t$ then $T_{n+1} = T_n$. If for some $t\in T_n$, $t$ extends
$s_{n+1}$ then $T_{n+1} = T_n \cup Z$ where $Z$ is defined as follows.
Let $k$ be the length of
the longest sequence in $T_n$. Let
\[Z = \{t\in\BINARY\MID \LENGTH{t} = k, s_{n+1}\subset t
\text{ and }\forall t' \in T_n, t'\not\subset t\}.\]
It is easy to see that the $T$ constructed in this way
satisfies the conditions of the
lemma. \QED

\noindent
The
following lemma is proved in \citet[p.\ 74]{Li}.
\begin{lem}\label{Kraft}{(\sc Kraft)}
Let $\ell_i$ be a sequence of natural numbers.
There is a \PF\ subset of \BINARY\
with this sequence as lengths of its members iff
\[
\sum 2^{-\ell_i} \le 1.
\]
\end{lem}

\noindent
Here is an effective version, proved in
\citet[Thm.\ 3.6.1, p.\ 125]{DH}.

\begin{lem}\label{kraftEf}
Let $\ell_i$ be a recursive enumeration of lengths such that
\[
\sum_i 2^{-\ell_i} \le 1.\] Then there is a recursive enumeration $a_i$
of a \PF\ set
such that $\LENGTH{a_i} = \ell_i$.
\end{lem}

\iffalse
The next lemma is easy
to verify by similar reasoning (namely, by recognizing that \PF\ subsets of
\BINARY\ correspond to non-overlapping paths through the binary tree).
\begin{lem}\label{genKraft}
Let $S\subseteq\BINARY$ be infinite and \PF. Then for every $n\in\N$,
\[
\sum \left\{2^{-\LENGTH{b}}\MID b\in S \mbox{ and } \LENGTH{b}\le n\right\} \le 1 - 2^{-n}.
\]
\end{lem}
\fi

\section{Prefix-free complexity}\label{Vsection}
Given a \PF\ \TM\ $L$ and $b\in\BINARY$, we let
$\PC_L(b)$ be the length of a shortest $a\in
\BINARY$ such that $L(a) = b$; in the absence of any such
$a$ $\PC_L(b) = \infty$.

\begin{defn}\label{pumdef} A \TM\ $V$ is \OB{\PF\ universal}
just in case $V$ is \PF\ and for every \PF\ \TM\
$L$ there is a constant $k\in\N$ such that for all
$b \in \BINARY$,
$\PC_V(b) \le \PC_L(b) + k$.
\end{defn}

\begin{prop}\label{pum}
\PFC\ universal \TM s exist.
\end{prop}

\PROOF It is easy to verify the existence of %a constant $k$ and
a uniform-effective procedure $P$ operating on \TM s such that
for all machines $M$:
\begin{enumerate}
\item $P(M)$ is \PF.
\item\label{pfa} $P(M)$ computes the same function as $M$, if $M$ is \PF.
%\item\label{pfc} $\LENGTH{P(M)} \le \LENGTH{M} + k$.
\end{enumerate}

\iffalse
Informally, given input $b$, $P(M)$ works by dovetailing
computations $M(a)$ for all $a\in \BINARY$. In these dovetailing
computations, let $c\in\BINARY$ be first such that
$M(c)\DOWARR$ and $c$ lies on the subtree of \BINARY\ that runs
through $b$; if there is no such $c$ then $P(M)(b)\UPARR$. If $c =
b$ then $P(M)(b) = M(c) = M(b)$; otherwise, $P(M)(b)\UPARR$.
Thus, $P(M)$ is not defined twice on any path through
\BINARY\ so $P(M)$ is \PF.
For (\ref{pfa}), if $M$ is \PF\ and converges on $b\in\BINARY$,
then $M(a)\UPARR$ for all $a\in \BINARY$ such that $a\subset b$
or $b\subset a$ (by the definition of \PF). Hence $P(M)(b) = M(b)$.
\fi

Recall from Section \ref{Notsec} our effective ordering $L_i$ of the
\TM s. Let \TM\ $V$ be such $V(1^\ell\CONCAT 0\CONCAT a) = P(L_\ell)(a)$ for
all $\ell\in\N$ and be undefined for inputs of any other form.
To see that $V$ is \PF, suppose that $a,b\in\BINARY$ were such that $a\subset b$
and both $V(a)\DOWARR$ and $V(b)\DOWARR$. Then for some $\ell\in\N$,
$a$ and $b$ have the forms $1^{\ell}\CONCAT 0\CONCAT c$ and
$1^{\ell}\CONCAT 0\CONCAT d$, respectively, with $c\subset d$. But then
$P(L_\ell)(c)\DOWARR$ and
$P(L_\ell)(d)\DOWARR$, contradicting the \PF\ nature of $P(L_\ell)$.

To finish the proof, let $\ell$ be the least index of
\PF\ \TM\ $L$. Then, for all $a,b\in\BINARY$, $L(a) = b$ implies
$V(1^\ell\CONCAT 0\CONCAT a) = P(L_\ell)(a) = L(a) = b$. Hence, for all
\PF\ \TM s $L$ with smallest index index $\ell$, $\PC_V(b) \le \PC_L(b)
+ \ell + 1$ for all $b\in\BINARY$. \QED

In light of the proposition, we fix a universal \PF\ \TM\ $V$, and we write
$\PC(b)$ in place of $\PC_V(b)$. It is clear that for every $b\in\BINARY$
there is a \PF\ \TM\ $L$ with $L(b) = b$. We infer immediately that:
\begin{lem}\label{evFact1x}
The range of $V$ is \BINARY.
%There is $m\in N$ such that for every $b\in\BINARY$,
%$\PC(b) \le \LENGTH{b} + m$.
\end{lem}
Hence, $\PC(b)$ is defined for all $b\in\BINARY$.
\section{Short \PC\ programs}

\begin{defn}\label{shortDefpc}
Call $p\in\BINARY$ \OB{\PF\ short} if $V(p)\DOWARR$\ and for all
$q\in\BINARY$, if $V(p) = V(q)$ then $\LENGTH{p} \le \LENGTH{q}$.
\end{defn}

\iffalse
\noindent Equivalently:
\begin{dispar}\label{shortEqpc}
$p\in\BINARY$ is \PF\ short iff
$\PC(V(p)) = \LENGTH{p}$.
\end{dispar}
\fi

\begin{prop}\label{shortFactpc}
There is no effective enumeration of an infinite number
of \PF\ short members of \BINARY.
\end{prop}

\PROOF
The argument is parallel to that for Proposition \ref{shortFact}.
Let $T = \{1^n0\MID n\in\N\}$. Then $T$ is \PF, infinite, and
effectively enumerable by increasing length, say, as $t_i$. So:
\begin{dispar}\label{shortpc1}
for all $m\in\N$, $\LENGTH{t_m} = m + 1$.
\end{dispar}
For a contradiction, let $p_i$, $i\in\N$ be an effective enumeration of infinitely
many \PF\ short members of \BINARY. Of course,
$\{\PC(p_i)\MID i\in\N\}$ is unbounded. So the following function
$\psi:T\rightarrow\BINARY$ is computable.
\begin{dispar}\label{shortpc2}
For all $m\in\N$, $\psi(t_m) = p_i$, where $i$ is least such that
$\PC(p_i) \ge 2m$.
\end{dispar}
By the definition of $\psi$:
\begin{dispar}\label{shortpc8}
for all $m\in\N$, $\PC(\psi(t_m))\ge 2m$.
\end{dispar}
Suppose that \TM\ $L$ computes $\psi$. Then $\DOMAIN(L) = T$ is
\PF, so by Definition \ref{pumdef} there is $k\in\N$ such that:
\begin{dispar}\label{shortpc9}
for all $m\in\N$, $\PC(\psi(t_m))\le \PC_L(\psi(t_m)) + k$.
\end{dispar}
Also, since $t_m$ causes $L$ to produce $\psi(t_m)$, and by
\ref{shortpc1}:
\begin{dispar}\label{shortpc10}
for all $m\in\N$, $\PC_L(\psi(t_m)) \le \LENGTH{t_m} = m + 1$.
\end{dispar}
{}From \ref{shortpc8} and \ref{shortpc9}:
\begin{dispar}\label{shortpc11}
for all $m\in\N$, $2m \le \PC_L(\psi(t_m)) + k$.
\end{dispar}
And from \ref{shortpc10} and \ref{shortpc11}:
\begin{quote}
for all $m\in\N$, $2m \le m + 1 + k$,
\end{quote}
which is false no matter which $k\in\N$ is chosen. \QED

\section{Subadditivity for \PC}

In contrast to Proposition \ref{subAdd1}, we have:

\begin{prop}\label{subAdd2} There is $k\in N$ such that
for all $a,b\in\BINARY$,
$\PC(a\CONCAT b) \le \PC(a) + \PC(b) + k$.
\end{prop}

As a preliminary to the proof, let $a,b,c,d,e\in\BINARY$ be
such that $b = a\CONCAT c$ and $b = d\CONCAT e$ with $a\neq d$ (hence,
$c \neq e$). Then no \PF\ \TM\ can be defined on both $a$ and $d$ because
one is a subsequence of the other. Since $V$ is \PF, we thus have:

\begin{lem}\label{subaddLem1}
For all $b\in\BINARY$ there is at most one pair $a,c\in\BINARY$ such that
$b = a\CONCAT c$, $V(a)\DOWARR$ and $V(c)\DOWARR$. Moreover, if such a pair
$a,c$ exists, it can be found effectively (via dovetailing).
\end{lem}

\SPOOF{Proposition \ref{subAdd2}}
By Lemma \ref{subaddLem1} let\TM\ $L$ be such that for all $b\in\BINARY$,
$L(b) = V(a)\CONCAT V(c)$ for the unique $a,c\in\BINARY$ such that
$b = a\CONCAT c$, $V(a)\DOWARR$, and $V(c)\DOWARR$; if no such $a,c$
exist then $L(b)\UPARR$. To show that $L$ is \PF, suppose that
$b,b'\in\BINARY$ were such that $b'\subset b$, $L(b)\DOWARR$, and
$L(b')\DOWARR$. Then there are $a,c$ and $a',c'$ such that
$b = a\CONCAT c$, $V(a)\DOWARR$, $V(c)\DOWARR$,
$b' = a'\CONCAT c'$, $V(a')\DOWARR$, $V(c')\DOWARR$, and either $a\subset a'$,
$a'\subset a$, $c\subset c'$ or $c'\subset c$.
But this implies that $V$ is not \PF, contradiction.
Since $L$ is \PF, by Definition \ref{pumdef} let $k\in\N$
be such that:
\begin{dispar}\label{hiDann1}
for all $c\in\BINARY$, $\PC(c) \le \PC_L(c) + k$.
\end{dispar}

Now let $a,b\in\BINARY$ be given. Let $p,q\in\BINARY$ have shortest
lengths such that $V(p) = a$ and $V(q) = b$, respectively. [That such $p,q$
exist follows from Lemma \ref{evFact1x}.] Then by the definition of $L$,
\begin{dispar}\label{hiDann2}
$L(p\CONCAT q) = V(p)\CONCAT V(q) = a\CONCAT b$.
\end{dispar}
By \ref{hiDann1},
$\PC(a\CONCAT b) \le \PC_L(a\CONCAT b) + k$. By \ref{hiDann2},
$\PC_L(a\CONCAT b) + k \le \LENGTH{p\CONCAT q} + k =
\LENGTH{p} + \LENGTH{q} + k$. And by the choice of $p,q$,
$\LENGTH{p} + \LENGTH{q} + k = \PC(a) + \PC(b) + k$. Therefore,
$\PC(a\CONCAT b) \le \PC(a) + \PC(b) + k$. \QED

\section{Chaitin's halting probability}

Recall that $V$ is our reference \PF\ universal \TM.
We define the \OB{halting probability}, $\Omega$, as follows.
\begin{defn}\label{OmegaDef}
\[
\Omega = \sum\left\{2^{-|b|} \MID b\in\BINARY \mbox{ and }V(b)\DOWARR\right\}.
\]
\end{defn}
By Lemma \ref{evFact1x},
$\Omega > 0$ since $\{b\in\BINARY\MID
V(b)\DOWARR\} \neq \emptyset$.
On the other hand, by Lemma \ref{Kraft},
$\Omega \le 1$ inasmuch as $\DOMAIN(V)$ is \PF.
So $\Omega$ may be conceived as a probability, namely, as the
chance of hitting a sequence in $\DOMAIN(V)$ by flipping a fair coin.
Note that the numerical value of $\Omega$ depends on the choice $V$ of
reference universal Turing Machine.

Define
$P_n = \{b\in\BINARY\MID  \LENGTH{b} \le n \mbox{ and } V(b)\DOWARR\}$.
Of course, for all $n\in\N$, $P_n$ is finite.
Following the development in \citet[p. 217]{Li} (but with some modifications),
we now establish:
\begin{lem}\label{chaitin3}
There is a computable function $\psi$ from $\BINARY$ to finite subsets
of \BINARY\ such that for all $n\in\N$, $\psi(\Omega[n]) = P_n$.
\end{lem}
\PROOF\
\iffalse
Recall that we
sometimes interpret $b\in\BINARY$ as the rational number
$\sum\{2^{-n}\MID n \le \LENGTH{b}\mbox{ and } b(n) = 1\}$.
By Lemma \ref{evFact1x}, $\DOMAIN(V)$ is infinite (and \PF). Therefore, by
Lemma \ref{genKraft}, for every $n\in\N$, $\sum\{2^{-\LENGTH{b}}\MID b\in P_n\}
\le 1 - 2^{-n}$.
\fi
%
First we demonstrate:
\begin{dispar}\label{Xbig}
Let $X \subseteq \DOMAIN(V)$ and suppose that
$\sum\{2^{-\LENGTH{b}}\MID b\in X\} > \Omega[n]$.
Then $P_n \subseteq X$.
\end{dispar}
For a contradiction,
suppose that $P_n \not\subseteq X$.
Then, for some $b \in \DOMAIN(V)$ with $\LENGTH{b} \le n,$ $b \not\in X.$
Since $X\subseteq\DOMAIN(V)$ it follows that:
\begin{dispar}\label{Xsmall}
$\sum\{2^{-\LENGTH{b}}\MID b\in X\} \le \Omega - 2^{-n}.$
\end{dispar}
But by \ref{geofacts}(\ref{chaitin2b}), $ \Omega - 2^{-n} \le \Omega[n],$ which with \ref{Xsmall}
contradicts the assumption $\sum\{2^{-\LENGTH{b}}\MID b\in X\} > \Omega[n]$, proving
\ref{Xbig}.

It was noted above that $\Omega > 0$ [indeed, Lemma \ref{evFact1x} implies
that $\DOMAIN(V)$ is infinite].
Therefore,
by \ref{geofacts}(\ref{oddconvention}) for every $n\in\N$,
$\Omega > \Omega[n]$. It follows at once that:
\begin{dispar}\label{OmegaInf}
For every $n\in\N$ there is a finite subset $X$ of $\DOMAIN(V)$
such that $\sum\{2^{-\LENGTH{b}}\MID b\in X\} > \Omega[n]$.
\end{dispar}

Now let us describe how to compute $\psi$.
Given $a\in\BINARY$, use dovetailing to enumerate $\DOMAIN(V)$.
Let $X\subseteq \DOMAIN(V)$ be the first finite subset that
emerges from the enumeration with the property that
$\sum\{2^{-\LENGTH{b}}\MID b\in X\} > a$. [If no such
$X$ is found then $\psi(a)\UPARR$.] Set
$\psi(a) = \{b\in X\MID \LENGTH{b}\le \LENGTH{a}\}$.

Let $n\in\N$ be given. To finish the proof of Lemma \ref{chaitin3},
we show that $\psi(\Omega[n]) = P_n$.
By \ref{OmegaInf}, the enumeration of $\DOMAIN(V)$
yields a finite subset $X$ such that
$\sum\{2^{-\LENGTH{b}}\MID b\in X\} > \Omega[n]$.
By \ref{Xbig}, $P_n\subseteq X$. Since $P_n$ contains the
members of $\DOMAIN(V)$ with length bounded by $n$,
$\psi(\Omega[n]) = \{b\in X\MID \LENGTH{b}\le \LENGTH{\Omega[n]}\} =
\{b\in X\MID \LENGTH{b}\le n\} = P_n$.
\QED

\begin{cor}\label{omegaCor}
$\Omega$ is not computable. That is, the function mapping
$n\in\N$ to $\Omega(n)$ is not effective.
\end{cor}

\PROOF\ Suppose for a contradiction that $\Omega$ is computable. Then
Lemma \ref{chaitin3} implies that
$P_n$ is computable from $n$.
The set $P = \{p\in\BINARY\MID V(p)\DOWARR\}$ is
therefore decidable. (Given $b\in\BINARY$, $b\in P$
iff $b\in P_{\LENGTH{b}}$.) We may therefore enumerate
$P$ in order of increasing length. Thus, we can effectively
enumerate the set $S$ of $p_j\in P$ such that for no $i < j$,
$V(p_i) = V(p_j)$. Each such $p_j$ is \PF\ short
in the sense of Definition \ref{shortDefpc}.
Since $\RANGE(V)$ is infinite, it is clear that
$S$ is infinite.
Such an enumeration is impossible
by Proposition \ref{shortFactpc}. \QED

Since all rational reals are computable, we also have:
\begin{cor}\label{omegaCor2}
$\Omega$ is irrational. In particular, $\Omega < 1$.
\end{cor}

Now we show the pivotal fact:

\begin{prop}\label{OmegadoesWork}
There is a constant $k$ such that
for all $n\in N$, $\PC(\Omega[n]) > n - k$.
\end{prop}

\PROOF\
{}From Lemma \ref{chaitin3} it follows that there is computable
$\varphi:\BINARY\rightarrow \BINARY$ such that for all $n\in\N$,
$\varphi(\Omega[n])\in\RANGE(V)$, and
for all $b\in\BINARY$, $\varphi(\Omega[n]) = V(b) \Rightarrow \LENGTH{b} > n$.
Informally, $\varphi$ is computed as follows.
Given $\Omega[n]$, compute
$P_n$ then compute $X = \{V(b)\MID b\in P_n\}$. Enumerate the range of
$V$ until the first $a\in\BINARY$ appears that is not in $X$. Set
$\varphi(\Omega[n]) = a$. Hence:
\begin{dispar}\label{OmegaWorksAx}
For all $n\in N$, $\PC(\varphi(\Omega[n])) > n$.
\end{dispar}
Let \TM\ $L$ be such that for all $b\in\BINARY$,
$L(b) = \varphi(V(b))$ with $L(b)\UPARR$ if $V(b)\UPARR$.
Then $L$ is \PF\ because $V$ is. Note that if $V(b) = \Omega[n]$ then
$L(b) = \varphi(\Omega[n])$. It follows that:
\begin{dispar}\label{xdan1}
For all $n\in\N$, $\PC_L(\varphi(\Omega[n])) \le \PC(\Omega[n])$.
\end{dispar}
Since $L$ is \PF, by Definition \ref{pumdef}, let $k\in\N$ be such that:
\begin{dispar}\label{xdan2}
For all $n\in\N$, $\PC(\varphi(\Omega[n])) \le \PC_L(\varphi(\Omega[n])) + k$.
\end{dispar}
It follows at once from \ref{xdan1} and \ref{xdan2} that:
\begin{dispar}\label{xdan3}
For all $n\in\N$, $\PC(\varphi(\Omega[n])) \le \PC(\Omega[n]) + k$.
\end{dispar}
{}From \ref{OmegaWorksAx} and \ref{xdan3} we obtain
\begin{quote}
For all $n\in\N$, $n < \PC(\Omega[n]) + k$.
\end{quote}
which implies Proposition \ref{OmegadoesWork}. \QED

\begin{defn}\label{randomKDef}
Any real $x$ for which there is a constant $k$ such that
for all $n\in N$, $\PC(x[n]) \ge n - k$ is
called \OB{random in the sense of Kolmogorov}.
\end{defn}
Proposition \ref{OmegadoesWork} thus shows:
\begin{cor}\label{xdan5}
There are reals
that are random in the sense of Kolmogorov.
\end{cor}
Later we'll see that the set of such reals has measure 1.

Notice that Definition \ref{pumdef} implies that the class
of reals that are random in the sense of Kolmogorov is invariant
under different choices of \PF\ universal \TM.

\section{\ML\ tests in sense 1}

With minor differences, we
start by following \cite[p. 141ff.]{Li}.
All probabilities in what follows
are with respect to the uniform product measure
(i.e., the coin flip measure). Given $S\subseteq\BINARY$,
let $O(S)$ be the set of reals that start with some
$\sigma\in S$. To reduce clutter, we rely on the following
convention.
\begin{conv}\label{probConv1}
Given $S\subseteq\BINARY$, we write
$\PROB(S)$ for $\PROB(O(S))$.
\end{conv}
The following lemma reflects the fact that members of a \PF\ subset
of \BINARY\ dominate non-intersecting neighborhoods of the Cantor Space.
\begin{lem}\label{useful1}
Let $X\subseteq\BINARY$ be \PF. Then $\PROB(X) = \sum_{b\in X}2^{-|b|}$.
\end{lem}

For a function $f$, we write
$f(x)\DOWARR = y$ to mean that $f(x)$ is defined and equals
$y$ (and similarly for inequalities).

\begin{defn}\label{smlt1}
Any partial recursive function $t:\BINARY\rightarrow N$ is called
a \OB{\ML\ test (in sense 1)} provided that for all $m \in N$,
\[
\PROB\{b\in\BINARY\MID t(b)\DOWARR\ge m\} \le 2^{-m}.\]
\end{defn}

\begin{ex}\label{smltex1}
Let $t:\BINARY\rightarrow N$ count the length of the
initial sequence of $0$'s in a given $x\in\BINARY$. Then
$t$ is a \ML\ test because for each $m\in N$, the set of
reals that begin with at least $m$ $0$'s has probability
$2^{-m}$. (For example, the probability of a real
beginning with three $0$'s is $1/8$.)
\end{ex}
To clarify notation, let us expand the foregoing example. Given
$m\in\N$, $S = \{b\in\BINARY\MID t(b)\DOWARR\ge m\}$ is a subset of
\BINARY. To calculate the probability of $S$, we consider the set of
reals $x$ that extend some member of $S$, that is, we consider the
set $O(S)$. The relevant condition is therefore that $\PROB(O(S)) \le
2^{-m}$. Relying on Convention \ref{probConv1}, the latter inequality
is written as $\PROB(S) \le 2^{-m}$.

\begin{ex}\label{smltex2}
Let $t:\BINARY\rightarrow N$ count the number of
consecutive even positions in a given $b\in\BINARY$ that
are filled with $1$'s starting from position $0$.
For $m = 3$, the probability of a real beginning with $1\_1\_1$
is $1/8$, and more generally, the probability of beginning
with at least $m$ $1$'s in even position is $2^{-m}$. So $t$ is a \ML\ test.
\end{ex}

\begin{ex}\label{smltex3}
Let $t:\BINARY\rightarrow N$ count the number of
times $101$ appears in a given binary sequence. Then $t$
is \textit{not} a \ML\ test. Indeed, for $m = 3$, the probability
of a real containing at least $m$ occurrences of $101$ is
unity (which exceeds $1/8$).
\end{ex}

\begin{ex}\label{smltex4a}
Let $t:\BINARY\rightarrow N$ count the number of
consecutive $0$'s just after the
initial segment $111$ if it occurs. Then $t$ is a \ML\ test even though
$t$ is partial. For each $m\in N$, the set of reals
that begin with at least $m$ $0$'s following the
initial sequence $111$ has probability $2^{-m+3}$.
\end{ex}

Since $t$ can be \textit{any} partial recursive function, the
collection of \ML\ tests captures all sufficiently rare patterns
that can be mechanically detected in binary sequences. The idea of
``sufficient rareness'' is given by the condition
\[
\PROB\{b\in\BINARY\MID t(b)\DOWARR \ge m\} \le 2^{-m}\]
in Definition \ref{smlt1}.

\begin{defn}\label{smlt2}
Let a \ML\ test $t$ and a real $x$ be given. We say that
$x$ \OB{passes} $t$ if $\{t(x[n])\MID n\in N\mbox{ and } t(x[n])\DOWARR\}$
is bounded. Otherwise, $x$ \OB{fails} $t$.
\end{defn}
The idea is that $x$ passes $t$ if $x$ doesn't
manifest ever more improbable events according to $t$
(namely, with probabilities declining as $2^{-m}$).

\begin{ex}\label{smltex4}
Let $t$ be as in Example \ref{smltex2}. Then a real $x$
passes $t$ if and only if $x(2n) = 0$ for some $n\in\N$.
\end{ex}

\begin{defn}\label{smlt3}
A real $x$ is \OB{\ML\ random (in sense 1)} just in case
$x$ passes every \ML\ test (in sense 1).
\end{defn}

\section{\ML\ tests in sense 2}

%Here we follow \cite[pp.\ 116 ff.]{Downey}, again with some variation.
\begin{defn}\label{imlt1}
Let function $f:N\rightarrow N$ be total recursive. Then $f$
is a \OB{\ML\ test (in sense 2)} provided
that for all $n\in N$,
$\PROB(W_{f(n)})\le 2^{-n}$.
\end{defn}
\begin{defn}\label{imlt2}
Let $f$ be a \ML\ test in sense 2.
A real $x$ \OB{passes} $f$ just in case
$x\not\in\bigcap\{O(W_{f(n)})\MID n\in\N\}$, and $x$ \OB{fails} $f$
otherwise).
We call $x$ \OB{\ML\ random (in sense 2)} if
$x$ passes all \ML\ tests (in sense 2).
\end{defn}

\begin{prop}\label{firstMLEqRL}
If a real is \ML\ random in sense 2 then it is \ML\ random
in sense 1.
\end{prop}

\PROOF\ Suppose that real $x$ fails \ML\ test $t$ in sense 1
[Definition \ref{smlt1}]. We must exhibit a \ML\ test $f$ in sense 2
[Definition \ref{imlt1}] that $x$ fails.

Let total
recursive $f:N\rightarrow N$ be such that $f(0)$ is an r.e.\ index for \BINARY, and for all $n > 0$,
\[W_{f(n)} = \bigcup\{t^{-1}(m) \MID m > n\}.\]Because $t$ is partial recursive, it is
clear that such an $f$ exists. To see that $f$ is a
\ML\ test in sense 2, suppose
for a contradiction that for some $n > 0$,
$\PROB(W_{f(n)}) > 2^{-n}$. Then
$\PROB(\bigcup\{t^{-1}(m) \MID m > n\}) > 2^{-n}$, hence:
\begin{dispar}\label{upshotx2}
\[\PROB\{b\in\BINARY\MID t(b)\DOWARR > n\} > 2^{-n}.\] % \mbox{ hence:}\]
\end{dispar}
But \ref{upshotx2} contradicts
the assumption that $t$ is a \ML\ test in sense 1 [Definition
\ref{smlt1}]. To show that $x$ fails
$f$, suppose otherwise. Since $x \in O(\BINARY) = O(W_{f(0)})$, there is $n > 0$ with \[x \not\in O(W_{f(n)}) =
O(\bigcup\{t^{-1}(m) \MID m > n\}) =
O(\{b\in\BINARY\MID t(b)\DOWARR \ge n+1\}).\]
But $x\not\in O(\{b\in\BINARY\MID t(b)\DOWARR \ge n+1\})$ implies that
$\{t(x[n])\MID n\in N\mbox{ and } t(x[n])\DOWARR\}$
is bounded, which contradicts the assumption that $x$
fails $t$ [Definition \ref{smlt2}]. \QED

\begin{prop}\label{firstMLEqLR}
If a real is \ML\ random in sense 1 then it is \ML\ random
in sense 2.
\end{prop}

\PROOF Suppose that real $x$ is not \ML\ random in sense 2. We will show
that $x$ is not \ML\ random in sense 1. Since $x$ is not \ML\ random in sense
2 there is total recursive $g:\N \rightarrow\N$ such that:
%\begin{dispar}\label{sp1}
\begin{enumerate}
\item\label{sp1a}
for all $n\in\N$, $\PROB(W_{g(n)}) \le 2^{-n}$
\item\label{sp1b} $x\in\bigcap\{O(W_{g(n)}\MID n\in\N)$.
\end{enumerate}
%\end{dispar}
Let total recursive $f:\N\rightarrow\N$ be such that
$f(n) = g(n+1)$. Then:
\begin{dispar}\label{sp2}
\begin{enumerate}
\item\label{sp2a}
for all $n\in\N$, $\PROB(W_{f(n)}) \le 2^{-(n+1)}$
\item\label{sp2b} $x\in\bigcap\{O(W_{f(n)}\MID n\in\N)$
%\item\label{sp2c} for all $n\in\N$, $W_{f(n)} \cap \{b\in\BINARY\MID \LENGTH{b} < n \} = \emptyset$.
\end{enumerate}
\end{dispar}
Define (possibly partial) recursive function $t:\BINARY \rightarrow \N$ such that for all
$b\in\BINARY$, $t(b) = \LENGTH{b} \text{ iff } b\in W_{f(|b|)}$, with $t(b)\UPARR$ if
$b\not\in W_{f(|b|)}$. To see that $t$ is a \ML\ test in sense 1, let $m\in\N$ be given.
Let $Z = \{b\in\BINARY\MID t(b)\DOWARR \ge m\}$. Then
$Z = \{b\in\BINARY\MID \LENGTH{b} \ge m\wedge b\in W_{f(|b|}\}\subseteq
W_{f(m)} \cup W_{f(m+1)} \dots$. So by \ref{sp2}\ref{sp2a} and Lemma \ref{geofacts}\ref{geofactsb},
\[\PROB(Z) \le \sum_{i = m+1}^{\infty}\left(\frac12\right)^{i} =
\left(\frac12\right)^{m}\] which  exhbits $t$
as a \ML\ test in sense 1. By \ref{sp2}\ref{sp2b},
$\{t(x[n])\MID n\in N\mbox{ and } t(x[n])\DOWARR\}$ is unbounded, hence
$x$ fails $t$ by Definition \ref{smlt2}. \QED

\begin{cor}\label{firstMLEqCor}
A real is \ML\ random in sense 1 if and only if
it is \ML\ random in sense 2.
\end{cor}

Henceforth we proceed in sense 2. That is:
\begin{conv}\label{sense2Conv}
By a \OB{test} is henceforth meant a \ML\ test in sense 2 [as described in
Definition \ref{imlt1}]. Likewise, a real is called \OB{\ML\ random} iff
it is \ML\ random in sense 2.
\end{conv}

\section{Universal tests}

\begin{defn}\label{UnivTestDef}
A test $f$ is \textit{universal} just in case for all reals
$x$, if $x$ fails any test then $x$ fails $f$.
\end{defn}

\begin{thm}\label{UnivTestThm}
There is a universal test.
\end{thm}

To prove the theorem, call tests $f$ and $g$ \textit{congruent} iff for
all $n\in \N$, $W_{f(n)} = W_{g(n)}$. Congruent
tests may not be identical since they might exploit
different indices for the same recursively enumerable set. Plainly:

\begin{lem}\label{UnivTestLem0}
Tests congruent to each other
are failed by the same set of reals.
\end{lem}

Let $\varphi_i$ be the usual indexing of partial
recursive functions.

\begin{lem}\label{UnivTestLem1}
There is total recursive $h:\N \rightarrow \N$ such that for all $i\in \N$:
\begin{enumerate}
\item\label{UnivTestLem1a}
for all $n\in \N$, if $\varphi_{h(i)}(n)\DOWARR$ then $\PROB(W_{\varphi_{h(i)}(n)}) \le 2^{-n}$;
and
\item\label{UnivTestLem1b} if $\varphi_i$ is a test then $\varphi_{h(i)}$ is a test that is congruent
to $\varphi_i$.
\end{enumerate}
\end{lem}

\SPOOF{Lemma \ref{UnivTestLem1}} Informally, here is how to compute $h$. Let $i$ be given.
Then $h(i)$ is an effectively constructed index for the \TM\ $L$ that behaves as follows.
Given $n\in N$, $L$ computes $\varphi_i(n)$. If $\varphi_i(n)\UPARR$ then $L(n)\UPARR$. Otherwise,
suppose that $\varphi_i(n) = m$. Then $L$
constructs an index for the longest initial segment of the
canonical enumeration of $W_m$ whose sum of probabilities remains bounded by $2^{-n}$.
Therefore, $\PROB(W_{\varphi_{h(i)}(n)}) \le 2^{-n}$. Now suppose that
$\varphi_i$ is a test and let $n\in\N$ be given. Then $\varphi_i(n)\DOWARR$ so $L(n)\DOWARR$.
Moreover, $\PROB(W_{\varphi_i(n)}) \le 2^{-n}$ so $W_{\varphi_{h(i)}(n)} =
W_{\varphi_i(n)}$. Hence
$\varphi_{h(i)}$ is a test, and congruent
to $\varphi_i$.\QED

\SPOOF{Theorem \ref{UnivTestThm}}
Let $h$ be as described in Lemma \ref{UnivTestLem1}. Given $i,n\in N$, define $X(i,n)$ to
be $W_{\varphi_{h(i)}(i+n)}$ if this set is defined, $= \emptyset$ otherwise.
\iffalse
It follows immediately that:
\begin{dispar}\label{finUp3}
For all $i\in\N$, if $\varphi_{h(i)}$ is a test then
$W_{\varphi_{h(i)}(i+n)}\subseteq \bigcup\{X(i,n)\MID i\in N\}$.
\end{dispar}
\fi
By Lemma \ref{UnivTestLem1}\ref{UnivTestLem1a}, $\PROB(X(i,n)) \le 2^{-(i+n)}$. Hence:
\begin{dispar}\label{keyUnivThm1}
For all $n\in N$, $\PROB\bigcup\{X(i,n)\MID i\in N\} \le \sum_{i=1}^{\infty}2^{-n-i} = 2^{-n}\sum_{i=1}^{\infty}2^{-i} = 2^{-n}$.
\end{dispar}
A universal test $f$ may now be defined as follows.
Given $n\in\N$, $f$ dovetails the enumerations of $W_{\varphi_{h(i)}(i+n)}$, $i\in\N$, to uniformly
effectively construct an r.e.\ index $f(n)$ for
$\bigcup\{X(i,n)\MID i\in N\}$.
Thus:
\begin{dispar}\label{finUpU2}
For all $n\in\N$, $W_{f(n)} = \bigcup\{X(i,n)\MID i\in N\}$.
\end{dispar}
By \ref{keyUnivThm1}, $f$ is a test.
To see that $f$ is universal, suppose that real $x$ fails test $g$.
We must show that $x$ fails $f$.
By Lemma \ref{UnivTestLem1}\ref{UnivTestLem1b},
let $i\in N$ be such that $g$ is congruent with $\varphi_{h(i)}$.
By Lemma \ref{UnivTestLem0},
$x$ fails $h(i)$,
that is:
\begin{dispar}\label{finUpU1}
 $x\in\bigcap \{O(W_{\varphi_{h(i)}(n)})\MID n\in\N\}$.
\end{dispar}
%Let $n\in N$ be given.
To show that $x\in\bigcap \{O(W_{f(n)})\MID n\in\N\}$, and thus complete
the proof, it
suffices to show that
$x\in O(W_{f(n)})$ for given $n\in\N$.
But by \ref{finUpU1}, $x\in O(W_{\varphi_{h(i)}(i+n)})$. So
\iffalse by \ref{finUp3},\fi
$x\in O(X(i,n))\subseteq O(\bigcup\{X(i,n)\MID i\in N\}) = O(W_{f(n)})$ by \ref{finUpU2}. \QED

\begin{cor}\label{measureC}
The probability of the set of \ML\ random reals is $1$.
\end{cor}
To prove the corollary, we rely on two lemmas the first
of which may be found in \citep[Thm.\ 3.17]{Oxtoby}.
\begin{lem}\label{OxFact}
Suppose that $A_i$ is a descending $\supseteq$-chain of measurable
sets of reals.
\iffalse\footnote{That is, for all $i$,
$A_i \supseteq A_{i+1}$. The subsets of reals that arise in this paper
are all measurable since they are unions of basic open sets, intersections
of such unions, or complements thereof.}\fi
Then
\[
\PROB(\bigcap_i A_i) = \lim_{i \rightarrow \infty}\ \PROB(A_i).
\]
\end{lem}

\begin{lem}\label{canonLem}
For every test $f$ there is a test $g$ such that
\begin{enumerate}
\item for all $i\in\N$, $W_{g(i)}\supseteq W_{g(i+1)}$, and
\item a real fails $g$ if and only it fails $f$.
\end{enumerate}
Moreover, an index for $g$ can be found uniform effectively
from an index for $f$.
\end{lem}

\PROOF It suffices to let $g(n)$ be an index for
$W_{f(0)} \cap \dots \cap W_{f(n)}$. \QED

\SPOOF{Corollary \ref{measureC}}
For all $i\in\N$, $O(W_i)$ is measurable since it is the union of
basic open sets [namely, $\bigcup\{O(b)\MID b\in W_i\}$].
By Theorem \ref{UnivTestThm}
and Lemma \ref{canonLem}, let $g$ be a universal test
such that $\{W_{g(n)}\MID n\in\N\}$ forms a $\supseteq$-descending
chain. By Definition \ref{UnivTestDef}, real $x$ is \ML-random iff $x$ passes $g$.
By Definition \ref{imlt2}, the set of reals that fail $g$
is $\bigcap\{O(W_{g(n)})\MID n\in\N\}$, whose probability
is $\lim_{i \rightarrow \infty}\ \PROB(O(W_{g(n)}))$ by
Lemma \ref{OxFact}. By Definition \ref{imlt1}, the latter limit
is zero. Hence,
the set of reals that pass $g$ has probability $1$.
\QED

\iffalse
\PROOF
By Theorem \ref{UnivTestThm}, let $f$ be a universal test, and
let $X = \{x\MID\mbox{$x$ is a real that fails } f\}$.
By Definitions \ref{imlt1} and \ref{imlt2}, the set of
\ML\ reals that fail a given test has probability $0$, hence:
\begin{dispar}\label{unicor1}
$\PROB(\overline X) = 1$.
\end{dispar}
Let $Y = \{x\MID\mbox{$x$ is a real that fails some test}\}$.
Since $f$ is universal, Definition \ref{UnivTestDef} implies that
$X = Y$ so $\overline Y = \overline X$. But $\overline Y$ is the set
of \ML\ random reals, and $\PROB(\overline Y) = 1$ by
\ref{unicor1}. \QED
\fi

\section{Equivalence of the two conceptions of randomness}

It will be shown in this section that a
real is random in the sense of Kolmogorov [Definition
\ref{randomKDef}] iff it is random in the sense of \ML\
[Definition \ref{imlt2}]. We abbreviate the two
senses of randomness to ``KC'' and ``ML'' (The ``C'' in
``KC'' stands for ``Chaitin'').

\begin{prop}\label{MLtoKC}
If a real is ML-random then it is KC-random.
\end{prop}

\noindent The proof follows \citet[\S6.2]{DH}. We start with a
lemma.

\begin{lem}\label{downeyLem}
Let TM $M$ have prefix-free domain. Fix $k\in\N$, and let $S = \{b\in\BINARY\MID
\PC_M(b)\le \LENGTH{b} - k\}$. Then $\PROB(S) \le 2^{-k}\PROB(\DOMAIN(M))$.
\end{lem}

\PROOF
For each $b\in S$ let $c_b\in\BINARY$ be such that $\LENGTH{c_b} \le \LENGTH{b} - k$
and $M(c_b) = b$. Then:
\begin{eqnarray*}
\PROB(S) & \le & \sum\{2^{-|b|}\MID b\in S\}\\& \le & \sum\{2^{-(|c_b|+k)}\MID b\in S\}\\
& = & 2^{-k}\sum\{2^{-|c_b|}\MID b\in S\}\\ & \le &
2^{-k}\sum\{2^{-|d|}\MID d\in\DOMAIN(M)\}\\ & = & 2^{-k}\PROB(\DOMAIN{M}).
\end{eqnarray*}
The first inequality follows from Lemma \ref{Kraft} because $S$ is a subset of the
\PF\ set $\DOMAIN(M)$. The second inequality comes from the choice of $c_b$. The third
inequality follows from $c_b\in\DOMAIN{M}$. The last equality relies on
Lemma \ref{useful1}. \QED

\SPOOF{Proposition \ref{MLtoKC}}
Let total recursive $f:\N\rightarrow\N$ be such that for all $k \in\N$,
$W_{f(k)} = \{ b \in \BINARY \MID \PC(b) \le \LENGTH{b} - k\}$. A dovetailing
construction shows that such an $f$ exists. By
Lemma \ref{downeyLem}, for all $k\in\N$, $\PROB(W_{f(k)}) \le
2^{-k}\PROB(\DOMAIN(W_{f(k)}))$.
Since $\DOMAIN(W_{f(k)})\subseteq \DOMAIN(V)$, and the latter set is \PF,
we have by Lemma \ref{useful1} that $\PROB(\DOMAIN(W_{f(k)}) \le 1$. Hence,
for all $k\in\N$, $\PROB(W_{f(k)}) \le 2^{-k}$ which exhibits $f$ as a
\ML\ test. Now suppose that real $x$ is ML-random. Then for some
$k\in\N$, $x\not\in O(W_{f(k)})$. Hence, for all $m\in\N$,
$\PC(x[m]) \ge \LENGTH{b} - k$ so $x$ is KC-random by
Definition \ref{randomKDef}.
\QED

Here is the converse to Proposition \ref{MLtoKC}:

\begin{prop}\label{KCtoML}
If a real is KC-random then it is ML-random.
\end{prop}

%\SPOOF{Proposition \ref{KCtoML}}
\PROOF\
We prove the contrapositive.
Suppose that real $x$ is not ML-random. Then there
is total recursive $f:N \rightarrow N$ such that $x\in\bigcap_n
W_{f(n)}$ and $\PROB(W_{f(n)}) \le 2^{-n}$. Hence by Lemma
\ref{irredundancy-lemma}, there is total recursive $g:N\rightarrow
N$ such that $x\in \bigcap_n W_{g(n)}$, and for all $n$, $W_{g(n)}$
is prefix-free, and $\PROB(W_{g(n)}) \le 2^{-n}$. So:
\begin{dispar}\label{g2fact}
\begin{enumerate}
\item\label{g2facta} for all $n\in N$, $W_{g(2n)}$ is prefix-free,
\item\label{g2factb} $x\in \bigcap_n W_{g(2n)}$, and
\item\label{g2factc} for all $n\in N$, $\PROB(W_{g(2n)})\ \le\ 2^{-2n}$.
\end{enumerate}
\end{dispar}
{}From \ref{g2fact}\ref{g2facta},\ref{g2factc} via Lemma \ref{useful1}:
\begin{dispar}\label{g3fact}
\[\sum_{b\in W_{g(2n)}} 2^{- |b|}\ \le\ 2^{-2n}.\]
\end{dispar}
We now show:
\begin{dispar}\label{gsmall}
\[\sum_{n\in N}\sum_{b\in W_{g(2n)}} 2^{n - |b|} \le 1.\]
\end{dispar}
To demonstrate \ref{gsmall}, observe that for all $n\in N$,
\[
\sum_{b\in W_{g(2n)}} 2^{n - |b|}\ = \sum_{b\in W_{g(2n)}} 2^n2^{- |b|}\ =\
2^n\sum_{b\in W_{g(2n)}} 2^{- |b|}\ \le\  2^n 2^{-2n}\ =\  2^{-n},\]
where the inequality follows from \ref{g3fact}.
Summing over $n$ yields \ref{gsmall}.

Returning to the proof of Proposition \ref{KCtoML}, let $(n_i, b_i, \ell_i)$ be
a repetition-free, recursive enumeration of all triples with $n_i\in\N$, $b_i\in W_{g(2n_i)}$,
and $\ell_i = \LENGTH{b_i} - n_i$.
From \ref{gsmall} and the definition of $\ell_i$ we infer:
\[
1 \ge \sum_{n\in N}\sum_{b\in W_{g(2n)}} 2^{n - |b|} =
\sum_{n\in N}\sum_{b\in W_{g(2n)}} 2^{-(|b| - n)} =
\sum_{i\in N}2^{-\ell_i}.
\]
%From \ref{gsmall} we infer that
%\begin{dispar}\label{nugsmall}
%$\sum_{i\in N}2^{-\ell_i} \le 1$.
%\end{dispar}
% We see from \ref{gsmall} that $\ell_i \ge 0$ for all $i\in\N$.
%Moreover, by \ref{gsmall}, $\sum_{i\in N}\ell_i \le 1$.
Hence Lemma \ref{kraftEf} implies that there is a recursive enumeration $a_i$ of a \PF\
subset of \BINARY\ such that $\LENGTH{a_i} = \ell_i$ for all $i\in\N$. It follows from the
two recursive enumerations that there is
partial recursive function $\psi$ with domain $\{a_i\MID i\in\N\}$
such that $\psi(a_i) = b_i$ for all $i\in\N$. Since $\{a_i\MID i\in\N\}$ is \PF, so is $\psi$.
Thus we have:
\begin{quote}
For all $n\in\N$ and $b\in W_{g(2n)}$ there is $i\in\N$ such that $\LENGTH{a_i} = \LENGTH{b} - n$ and
$\psi(a_i) = b$.
\end{quote}
It follows immediately that:
\begin{dispar}\label{getc0}
For all $n\in\N$ and $b\in W_{g(2n)}$, $\PC_{\psi}(b) \le \LENGTH{b} - n$.
\end{dispar}
By Proposition \ref{pum}, since $\psi$ is \PF, choose $c$ such that for all
$b\in \BINARY$, $\PC(b) \le \PC_\psi(b) + c$. Then \ref{getc0} implies:
\begin{quote}
For all $n\in\N$ and $b\in W_{g(2n)}$, $\PC(b) \le \PC_{\psi}(b) + c \le \LENGTH{b} + c - n $.
\end{quote}
Substituting for $n$ in the foregoing, we obtain:
\begin{dispar}\label{getc1}
For all $k\in\N$ and $b\in W_{g(2(c+k))}$,
$\PC(b) \le \LENGTH{b} + c - (c + k) = \LENGTH{b} - k$.
\end{dispar}
In view of \ref{g2fact}\ref{g2factb},
for all $k\in\N$ there is $m\in \N$ such that
$x[m]\in W_{g(2(c+k))}$. So \ref{getc1} implies:
\begin{quote}
For all $k\in\N$ there is $m\in\N$ such that
$\PC(x[m]) \le \LENGTH{x[m]} - k = m - k$.
\end{quote}
By Definition \ref{randomKDef}, the last inequality shows $x$ not to
be KC-random. \QED

\begin{cor}\label{equivCor}
A real is ML-random if and only if it is CK-random.\end{cor}

\noindent
From the preceding corollary and Corollary \ref{measureC}:

\begin{cor}\label{KCMeasure}
The set of reals
that are random in the sense of Kolmogorov has probability $1$.
\end{cor}

\renewcommand{\baselinestretch}{1.0}
\bibliography{cohbib2}
\bibliographystyle{econometrica}
\end{document}